\documentclass[twocolumn,amsmath,amssymb,floatfix,pre,floatfix]{revtex4} 

\usepackage{graphicx,color}
\definecolor{brown}{rgb}{0.63,0.27,0.18}
\definecolor{orange}{rgb}{1.00,0.65,0.00}
\usepackage[dvipsnames]{xcolor} 

\usepackage{moresize}
\usepackage{dcolumn}
\usepackage{bm,multirow}

\begin{document}

\newcommand {\rsq}[1]{\langle R^2 (#1)\rangle}
\newcommand {\rsqL}{\langle R^2 (L) \rangle}
\newcommand {\rsqbp}{\langle R^2 (N_{bp}) \rangle}
\newcommand {\Nbp}{N_{bp}}
\newcommand {\etal}{{\em et al.}}
\newcommand{\Ham}{{\cal H}}
\newcommand{\ar}[1]{\textcolor{red}{#1}}
\newcommand{\dmi}[1]{\textcolor{ForestGreen}{#1}}
\newcommand{\scs}{\ssmall}



\title{Glassiness and Heterogeneous Dynamics in Dense Solutions of Ring Polymers}

\author{Davide Michieletto$^{1,*,\dagger}$, Negar Nahali$^{2,*}$ and Angelo Rosa$^{2,\dagger}$}
\affiliation{$^1$\mbox{School of Physics and Astronomy, University of Edinburgh, Peter Guthrie Tait Road, Edinburgh EH9 3FD, Scotland, UK}\\ $^2$\mbox{SISSA - Scuola Internazionale Superiore di Studi Avanzati, Via Bonomea 265, 34136 Trieste, Italy} \\$^*$Joint first author $^\dagger$For correspondence: D. Michieletto (davide.michieletto@ed.ac.uk), A. Rosa (anrosa@sissa.it)}


\begin{abstract}
Understanding how topological constraints affect the dynamics of polymers in solution is at the basis of any polymer theory and it is particularly needed for melts of rings.
These polymers fold as crumpled and space-filling objects and, yet, they display a large number of 
topological constraints.
To understand their role, here we systematically probe the response of solutions of rings at various densities to ``random pinning'' perturbations.
We show that these perturbations trigger non-Gaussian and heterogeneous dynamics, eventually leading to non-ergodic and glassy behaviours. We then derive universal scaling relations for the values of solution density and polymer length marking the onset of vitrification in unperturbed solutions. Finally, we directly connect the heterogeneous dynamics of the rings with their spatial organisation and mutual interpenetration.
Our results suggest that deviations from the typical behaviours observed in systems of linear polymers may originate from architecture-specific (threading) topological constraints.
\end{abstract}

\maketitle

\paragraph*{Introduction --} 

The behaviour of unknotted and mutually unlinked ring polymers in dense solutions and melts is a yet unsolved issue in Polymer Physics~\cite{McLeish2008},
and it has stimulated much  theoretical~\cite{KhokhlovNechaev85,grosbergJPhysFrance1988,ORD_PRL1994,mullerPRE2000,DeguchiJCP2009,Halverson2011_1,Halverson2011_2,RosaMicheletti2011,SakauePRL2012,RosaEveraersPRL2014,GrosbergSoftMatter2014,CaponeLikosSoftMatter2014,MichielettoTurner2014,SmrekGrosbergRingDyn2015,Lee2015,MichielettoTurner2016,PanyukovRubinsteinMacromolecules2016}
and 
experimental~\cite{kapnistos2008,RubinsteinSolvQualRings2015,Bras2011,Bras2014,Goossen2014,Vlassopoulos2016} work  in last decades. 
One of the most elusive aspects of this problem is the interplay between topological constraints (TCs) and both, structure and dynamics of the rings,
which looks far more intricate than in their linear analogs.
In the latter case, TCs induce slow dynamics through the reptative motion of the chain ends~\cite{deGennes71,DoiEdwards,RubinsteinColby} without affecting the average chain size or gyration radius, $R_g$, which remains essentially random-walk-like~\cite{Flory1969,degennes} and scales with the polymerization index $N$ as $R_g \sim N^\nu$ with $\nu=1/2$.
In the former case, the polymers have no ends to ``reptate''~\cite{McLeish2008} and global topological invariance requires that all rings remain permanently unlinked 
at the expense of some entropic loss~\cite{Cates1986}.
As a result, TCs affect both, dynamics and conformations of the rings whose gyration radius is characterized by a non-trivial exponent predicted to be in the range between $\nu=1/4$~\cite{KhokhlovNechaev85} and $\nu=2/5$~\cite{Cates1986}.

In recent years, more accurate computational work~\cite{DeguchiJCP2009,Halverson2011_1,RosaEveraersPRL2014} has provided evidence that in the limit of large $N$, $\nu \rightarrow 1/3$, in agreement with a picture in which rings fold into ``crumpled-globule''-like conformations~\cite{grosbergEPL1993}
whose compaction increases with solution density~\cite{NahaliRosa2016}.
In spite of this, the surface of each ring, {\it i.e.} the fraction of contour length in contact with other chains, is ``rough''~\cite{NahaliRosa2016} and scaling as $N^\beta$ with $\beta \lesssim 1$~\cite{Halverson2011_1,Halverson_JPA2013,GrosbergSoftMatter2014,Smrek2016minsurf}.
In fact, crumpled rings do not fully segregate or expel neighbouring chains from the occupied space~\cite{Halverson2011_1}, rather, they fold into interpenetrating or ``threading'' conformations~\cite{MichielettoTurner2014,Michieletto2014selfthreading} that are akin to interacting ``lattice animals''~\cite{RosaEveraersPRL2014} with long-range (loose) loops~\cite{PanyukovRubinsteinMacromolecules2016,Michieletto2016SoftMatter}.

\begin{figure}[t!]
	\centering
	\includegraphics[width=0.48\textwidth]{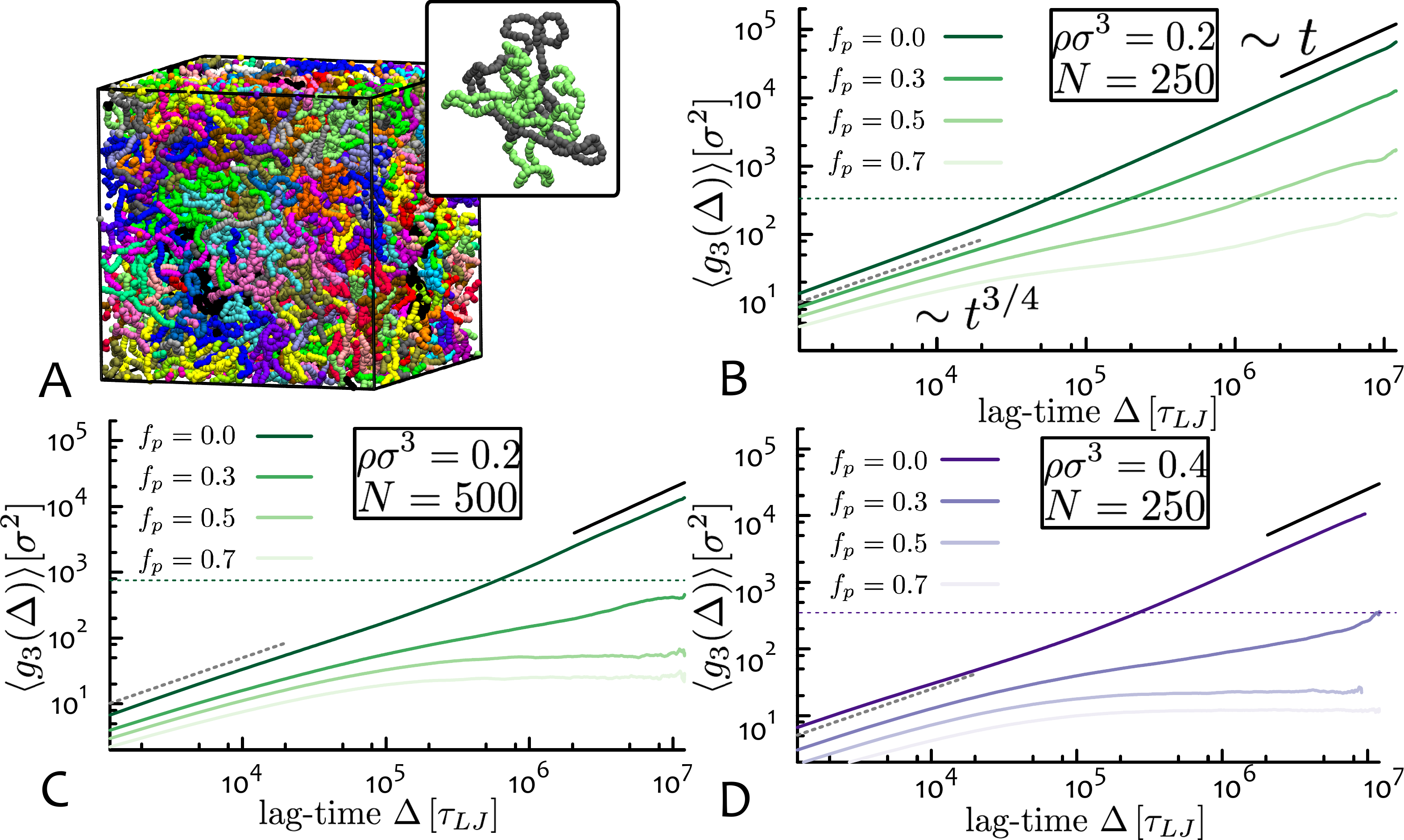}
	\vspace*{-0.5 cm}
	\caption{
		\textbf{Random Pinning Triggers Slowing Down and Glassiness.}		
		(\textbf{A})
		Typical melt structure for rings of $N=250$ monomers with $f_p=0$ and $\rho=0.2\sigma^{-3}$.
		Inset: Two rings isolated from the melt and showing mutual threading.
		(\textbf{B,C,D})
		Mean-square displacement of rings centre of mass, $\langle g_3(\Delta) \rangle$ (Eq.~(\ref{eq:EnsembleAvMSD})) as a function of lag-time $\Delta$ for ring solutions with
		selected $N$ and $\rho$.
		Rings display glassy behaviour (suppressed diffusion, $\langle g_3(\Delta) \rangle \sim \Delta^0$) for $f_p > f_p^\dagger$ where
		$f_p^\dagger$ is found to decrease with both, $N$ and $\rho$. 
		Dashed horizontal lines are for the mean-square ring diameter, $4 \langle R_g^2\rangle$. 
	}
	\vspace*{-0.5 cm}
	\label{fig:g3sRho03}
\end{figure}

Threadings are architecture-specific TCs that uniquely characterize systems of polymers whose contours display (quenched) closed loops (see Fig.~\ref{fig:g3sRho03}(A)).
By exploiting the abundance of these peculiar interactions,
it has been shown recently~\cite{MichielettoTurner2016} that 
a novel ``topological freezing'' can be induced in rings solutions at any temperature $T$. 
This putative glassy state, inherently driven by the topology of the constituents, 
is achieved by randomly pinning a fraction of rings, $f_p$,
above an empirical ``critical'' value (see Fig.~S\ref{fig:phasediag}(A) in Supplemental Material (SM)):
\begin{equation}
f_p^\dagger (N) = - f_N\log{\left(\dfrac{N}{N_g} \right)} \, ,
\label{eq:criticalfN}
\end{equation} 
where $N_g$ is the theoretical length required for spontaneous ({\it i.e.}, $f_p \rightarrow 0$) vitrification and $f_N$ a non-universal
parameter~\cite{PhaseDiagFigNote}.

Topological freezing is the consequence of the proliferation of inter-ring constraints~\cite{MichielettoTurner2014,MichielettoTurner2016},
with the latter depending either on the polymerization index, $N$, or the density of the solution, $\rho$.
While it has been shown that longer rings generate more TCs~\cite{MichielettoTurner2016}, it remains unclear how they behave if solutions become denser, rings more crumpled~\cite{NahaliRosa2016} and less space is available to threading. 

Motivated by these considerations, in this Letter we study the effect of TCs by ``randomly pinning'' solutions of semi-flexible ring polymers,
and probe the dynamic response of the rings for different solution densities and chain lengths. 
We show that the threshold pinning fraction $f_p^\dagger$ obeys an empirical relation akin to Eq.~\eqref{eq:criticalfN} and we derive universal scaling relations for the values of $N_g$ and $\rho_g$ at which spontaneous ($f_p \rightarrow 0$) glassiness is expected.
We further discuss the dynamics of rings in terms of ensemble- and time-average observables and report, for the first time, numerical evidence for ergodicity breaking effects and pronounced heterogeneous non-Gaussian dynamics, even in unperturbed ($f_p = 0$) solutions.

\paragraph*{Results --} 

We present the results of large-scale molecular dynamics (MD) simulations of solutions of semi-flexible ring polymers
made of $N=250$ and $N=500$ beads, for monomer densities $\rho\sigma^3 = 0.1, 0.2, 0.3, 0.4$ (see SM for details). 
For each combination of $N$ and $\rho$, we run a single, independent realisation of the system in which a random fraction $f_p$ of rings have been ``pinned'', {\it i.e.} permanently frozen in space and time.

The dynamics of a single non-frozen ring can then be captured
by the mean-square displacement of its centre of mass, $g_3(T,\Delta)$, as a function of the lag-time $\Delta$ and measurement time $T$:
\begin{equation}\label{eq:SingleRingMSD}
	g_3(T,\Delta) \equiv \dfrac{1}{T-\Delta} \int_0^{T-\Delta} \left[\bm{r}_{CM}(t+\Delta) -  \bm{r}_{CM}(t)\right]^2 dt \, .
\end{equation}
The time-average displacement can be defined as $g_3(\Delta) \equiv \overline{g_3(T, \Delta)}$ while its ensemble average as
\begin{equation}\label{eq:EnsembleAvMSD}
	 \langle g_3(T,\Delta) \rangle  \equiv \dfrac{1}{M_f}\sideset{}{'}\sum g_3(T,\Delta) \, ,
\end{equation}
with $\sum^\prime$ indicating that the average is performed over the set of $M_f$ ``free'', {\it i.e.} not explicitly pinned, rings.
Accordingly, we indicate the time- {\it and} ensemble-average displacement as $\langle g_3(\Delta) \rangle$.

Fig.~\ref{fig:g3sRho03}(B,C,D) directly compare the behaviour of $\langle g_3(\Delta)\rangle$ in response to the random pinning of different fractions $f_p$ of rings (see also SM Fig.~S\ref{fig:g3sRhoOthers} for more cases).
For unperturbed solutions ($f_p=0$), the data reproduce the known~\cite{Halverson2011_2,Bras2014,GrosbergSoftMatter2014} crossover from sub-diffusive ($\langle g_3(\Delta) \rangle \sim \Delta^{3/4}$) to diffusive ($\langle g_3(\Delta) \rangle \sim \Delta$) behaviour.
Perturbed systems, instead, display a reduced average diffusion, the more severe the higher the value of $f_p$.
In particular, for $f_p$ larger than $f_p^\dagger(\rho,N)$, the average displacement remains well below one ring diameter (marked by the horizontal dashed lines)
and does not diverge in time,
indicating~\cite{MichielettoTurner2016} a solid-like (glassy) behaviour. 
Furthermore, we find that $f_p^\dagger(\rho,N)$ decreases as a function of both, ring length $N$~\cite{MichielettoTurner2016} and, unexpectedly, monomer density $\rho$.

\begin{figure}[b!]
	\centering
	\includegraphics[width=0.48\textwidth]{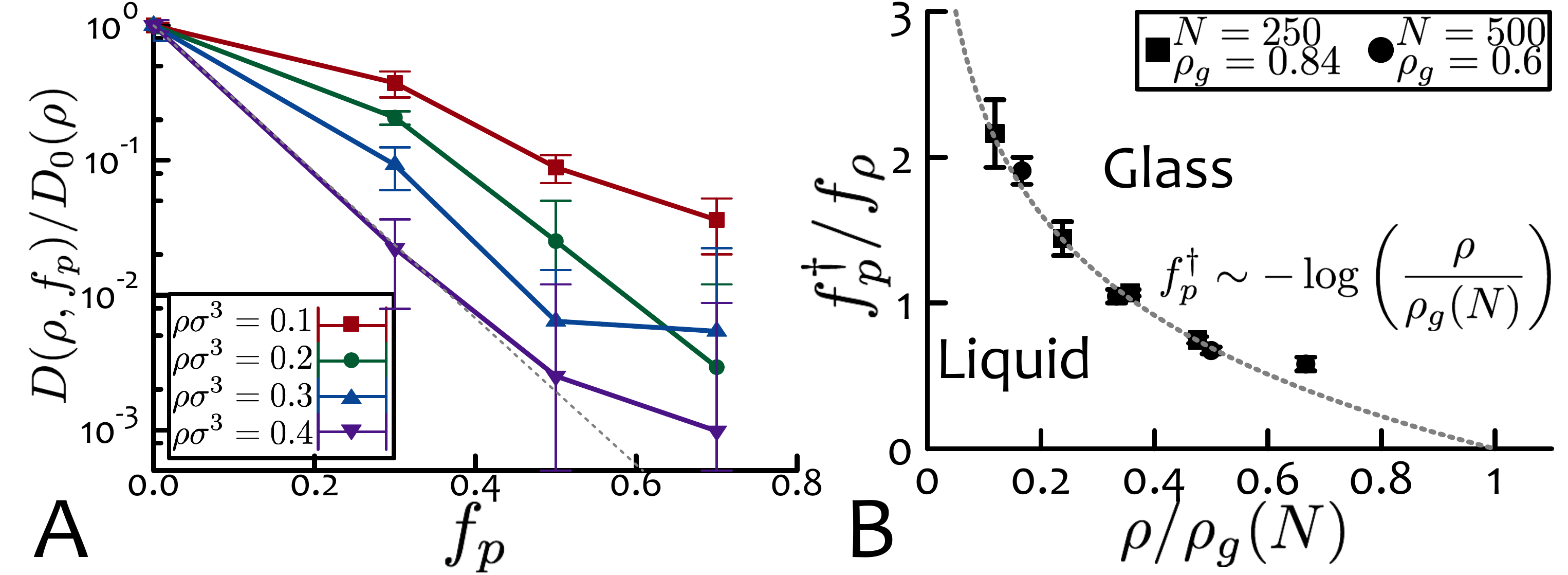}
	\vspace*{-0.8 cm}
	\caption{
		\textbf{Exponential Slowing Down and Universal Phase Diagram.}
		(\textbf A)
		$D(\rho, f_p)/D_0(\rho)$ is compatible with exponential decay (dashed line) in $f_p$.
		An arbitrarily small ($0.01$) value is chosen to determine the transition to glassy behaviour~\cite{MichielettoTurner2016}.
		(\textbf{B})
		Curve for $f_p^\dagger(\rho,N)/f_\rho$ as a function of $\rho/\rho_g(N)$ (see Eq.~\eqref{eq:criticalfrho}) showing collapse onto the universal curve $-\log{(x)}$ for $N=250$ and $N=500$ data. 
	}
	\label{fig:phasediag_scaling}
	\vspace*{-0.4 cm}
\end{figure}

In order to obtain the functional form of $f_p^\dagger(\rho, N)$,
the asymptotic diffusion coefficient $D(\rho, f_p) \equiv \lim_{\Delta\rightarrow\infty} \langle g_3(\Delta) \rangle/6\Delta$ at given ($N$, $\rho$, $f_p$)
is computed by best fit of the long-time behaviour of the corresponding $\langle g_3(\Delta) \rangle$ to a linear function (for details, see SM).
Fig.~\ref{fig:phasediag_scaling}(A) ($N=250$) and SM Fig.~S\ref{fig:phasediag}(B) ($N=500$) show plots for $D(\rho,f_p)/D_0(\rho)$ -- where $D_0(\rho)\equiv D(\rho,f_p=0)$ -- as a function of $f_p$.
Corresponding datasets are well fitted by the exponential function $d(f_p) = \exp{(-k f_p)}$, in agreement with previous results~\cite{MichielettoTurner2016}.
We thus extract $f_p^\dagger(\rho,N)$ by finding the intersection of $d(f_p)$ with a convenient small  value of $0.01$~\cite{MichielettoTurner2016}. 
The obtained ``critical'' lines $f_p^\dagger(\rho,N)$ (see SM Fig.~S\ref{fig:phasediag}(C)) separate regions of the parameter space $(\rho,f_p)$ with finite (liquid) and vanishing (glassy) diffusion coefficients.

Interestingly, we find that the functional form of $f_p^\dagger(\rho)$ is akin to Eq.~\eqref{eq:criticalfN}, {\it i.e.}
\begin{equation}
f_p^\dagger(\rho,N) = -f_\rho\log{\left(\dfrac{\rho}{\rho_g}\right)}\, ,
\label{eq:criticalfrho}
\end{equation} 
and that our data for $N=250$ and $N=500$ collapse onto a master curve $f_p^\dagger(x=\rho/\rho_g(N))/f_\rho=-\log{\left( x \right)}$ 
with $f_\rho=0.43$ (Fig.~\ref{fig:phasediag_scaling}(B)). 
Given that both, Eqs.~\eqref{eq:criticalfN} and \eqref{eq:criticalfrho}, describe 
the same quantity,
one may argue that the right-hand-side of both equations must be equal.
By combining them~\cite{MasterCurveNote} under the assumption 
that the only dependence on $\rho$ is contained in $N_g$,
the values of $\rho_g$ and $N_g$ for spontaneous {\it topological} vitrification obey the following universal scaling relations
\begin{align}\label{eq:scaling}
&\rho_g(N) \sim N^{-\eta} \, \notag\\
&N_g(\rho) \sim \rho^{-1/\eta} \, ,
\end{align}
with $\eta = f_N/f_\rho \simeq 0.7$ (using $f_N=0.303$ and $f_\rho=0.43$).
Eqs.~\eqref{eq:criticalfrho} and \eqref{eq:scaling} provide quantitative predictions that may readily be tested in 
computer simulations and future experiments on melts of rings. 

\begin{figure}[b!]
	\centering
	\includegraphics[width=0.48\textwidth]{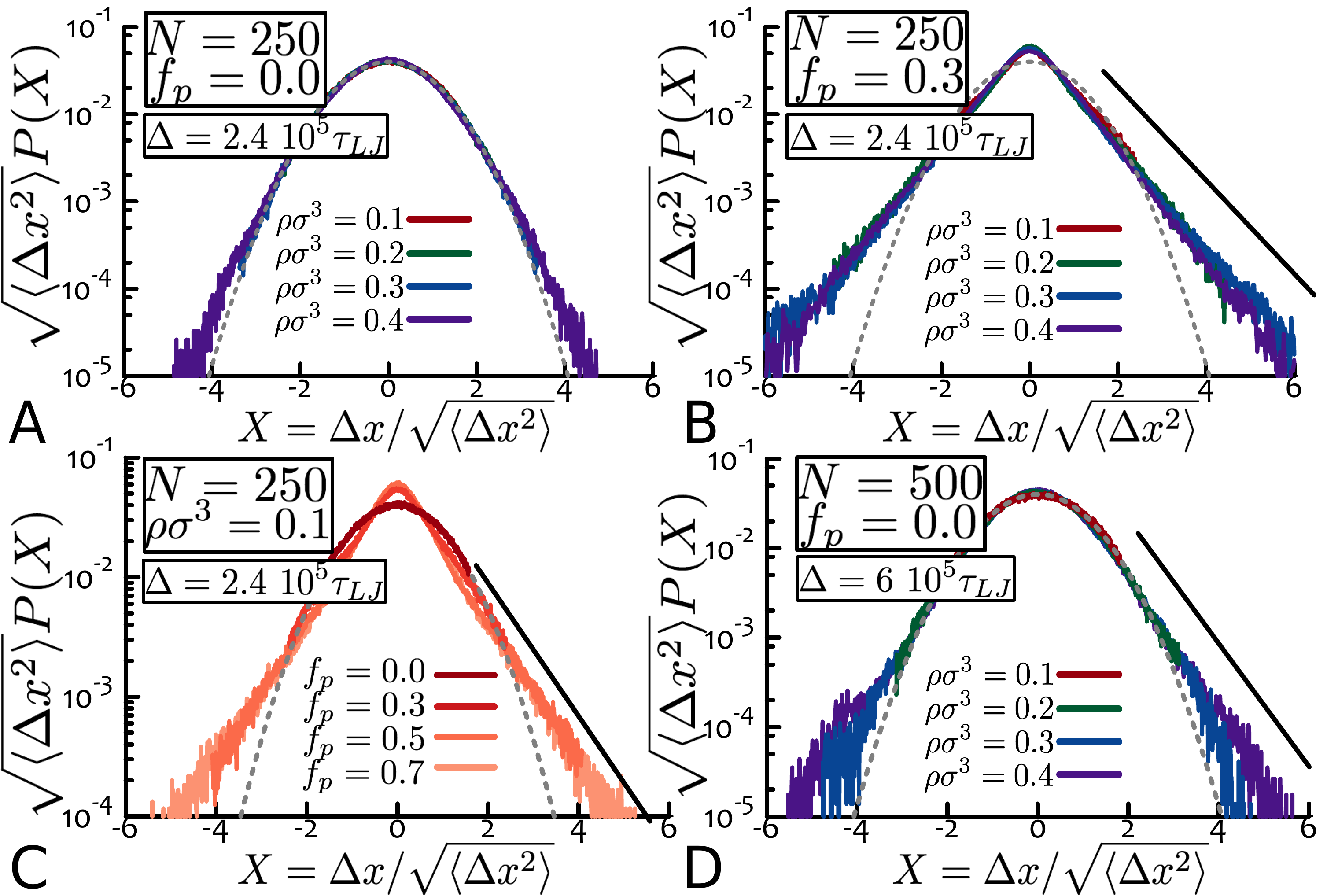}
	\vspace*{-0.4 cm}
	\caption{
		\textbf{Distributions of Displacements are non-Gaussian.} 		
		Distribution functions, $P(X)$, of $1d$ scaled displacements of the centers of mass of non-pinned rings, $X \equiv \Delta x/\sqrt{\langle \Delta x^2\rangle}$, at lag-times $\Delta$.
		$P(X)$ is described by a Gaussian function with zero mean and unit variance (dashed lines) in non-pinned systems (\textbf{A}), while it displays caging and fat, exponential tails (solid lines) in pinned solutions (\textbf{B},\textbf{C}).
		(\textbf{D}) Deviations from Gaussian behaviour (exponential tails) are also observed in unperturbed solutions with $N=500$ and $\rho \sigma^3\geq 0.3$.
	}
	\label{fig:Pdisplacements}
	\vspace*{-0.4 cm}
\end{figure}

Having determined the empirical functional form of $f_p^\dagger$ and the generic behaviours of $N_g$ and $\rho_g$,
we now turn our attention on the role of TCs in the dynamics of single rings.
To this end, we consider the distribution of $1d$ displacements $\Delta x$~\cite{NoteOnPDeltaX}
\begin{equation}
P(\Delta x)= \langle \delta(\Delta x - \left| x(t+\Delta)-x(t)\right|)\rangle \, ,
\end{equation}
which corresponds to the self-part of the van-Hove function~\cite{Kob1997,BerthierKobPRL2007} at given lag-time $\Delta$.
The distribution of rescaled displacements $X \equiv \Delta x/\sqrt{\langle \Delta x^2\rangle}$ is expected to be described by the universal Gaussian function with zero mean and unit variance~\cite{Kob1997}.
This is indeed the case for unperturbed solutions of short rings ($f_p=0$, Fig.~\ref{fig:Pdisplacements}(A)),
whereas both, perturbed solutions (Fig.~\ref{fig:Pdisplacements}(B,C)) and even unperturbed ones with long rings and high density (Fig.~\ref{fig:Pdisplacements}(D)),
distinctly deviate from the Gaussian behaviour (see also SM Figs.~S\ref{fig:PDisplacementsA}-S\ref{fig:PDisplacementsC}). 

Two novel features emerge from these plots:
First, a prominence of rings with short ``cage-like'' displacements, identified by the narrow region centred around $X=0$ where $P(X)$ remains above the Gaussian. Second, the appearance of a population of rings travelling faster than the average ring, giving rise to ``fat'' exponential tails.
Both features are akin to those observed in generic systems of particles close to glass and jamming transitions~\cite{BerthierKobPRL2007}
where each component alternates cage-like motion and large-scale rearrangements. 

The non-Gaussian behaviour detected in our systems is clearly triggered by pinning perturbations (Fig.~\ref{fig:Pdisplacements}(C)), arguably through threading TCs.
We conjecture that threading configurations may also be at the basis of the (weaker) non-Gaussian behaviour observed in unperturbed ($f_p=0$) solutions at large densities and $N=500$ (Fig.~\ref{fig:Pdisplacements}(D)).
In particular, threadings may in general be the reason of the cage-like, non-Gaussian, motion of large ring polymers seen in experiments~\cite{Bras2014}.
Although not permanent as in the case of pinned solutions, threadings between rings may in fact act as transient cages,
and our results suggest that they may be the more long-lived the denser the solutions and the longer the rings.  

\begin{figure}[t!]
	\centering
	\includegraphics[width=0.48\textwidth]{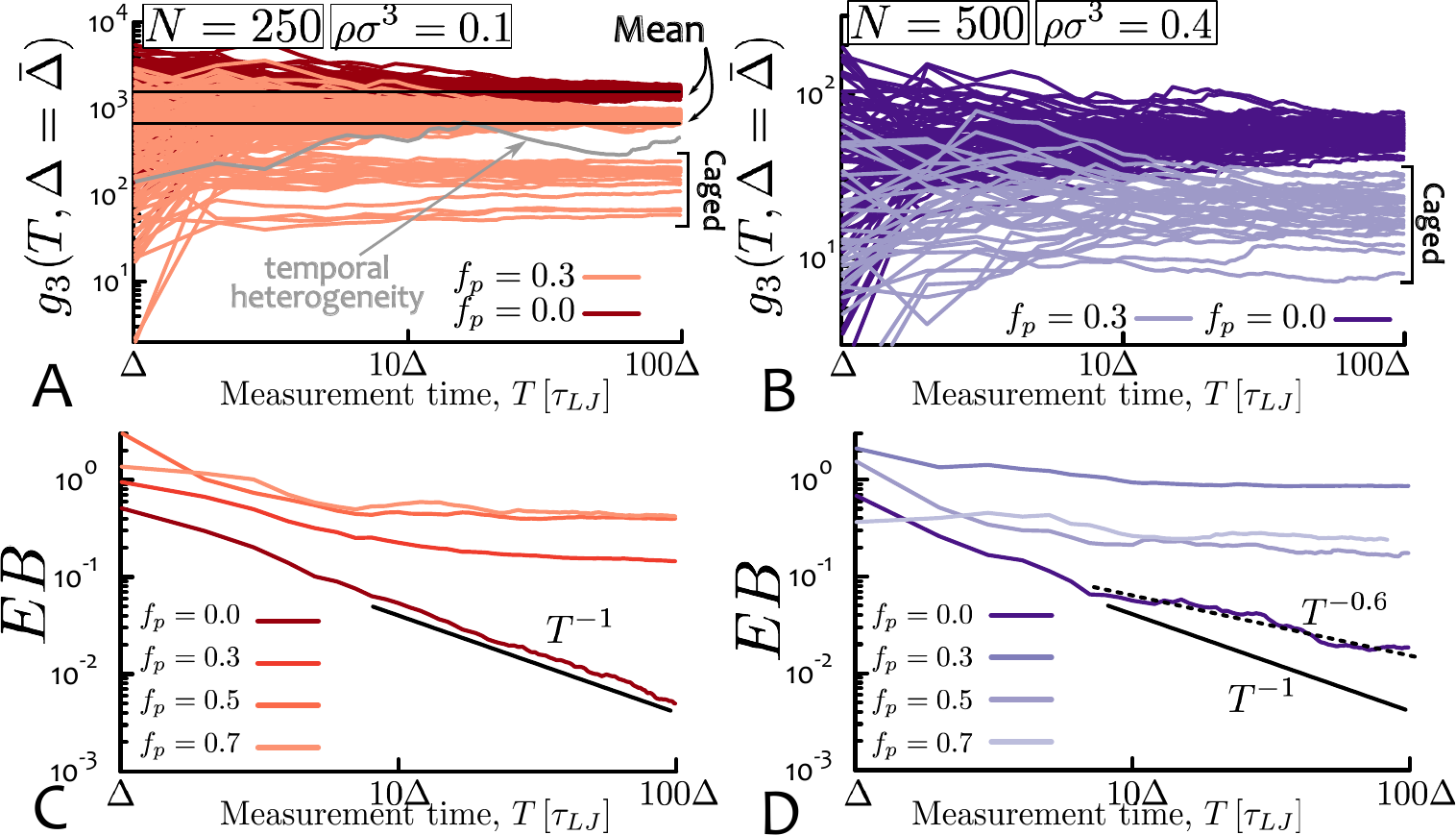}
	\caption{
		\textbf{Heterogeneous Dynamics and Ergodicity Breaking.}
		(\textbf{A,B})
		Single curves for $g_3(T,\overline{\Delta})$ at fixed lag-time $\Delta=\overline{\Delta}=1.2 \text{ }10^5$ $\tau_{\rm LJ}$ as a function of measurement time $T$.
		Representative results for $\rho\sigma^3=0.1$ and $N=250$ and $\rho\sigma^3=0.4$ and $N=500$. Thick horizontal lines represent the mean $\langle g_3(\overline{\Delta}) \rangle$. 
		(\textbf{C,D})
		Plots for the ergodicity-breaking ($EB$) parameter (Eq.~\eqref{eq:ErgBreakParam}).
		A decay $\sim T^{-1}$ is expected for standard diffusive processes, whereas $\sim T^0$ indicates strong ergodicity-breaking.
		Rings with $N=500$ at the highest density $\rho\sigma^3 = 0.4$ display non-Gaussian decay even at $f_p=0$.
	}
	\vspace*{-0.5 cm}
	\label{fig:Het}
\end{figure}

In order to better understand the observed deviations from Gaussian behaviour, we now investigate time-average quantities of single ring trajectories. In Fig.~\ref{fig:Het}(A,B) we report  $g_3(T,\overline{\Delta})$, {\it i.e.} the centre of mass displacement of single rings at fixed lag-time $\Delta=\overline{\Delta}$ and increasing measurement time $T$ (see also SM, Figs.~S\ref{fig:AgeingN250}-S\ref{fig:AgeingN500}).
Unperturbed solutions of short rings show that $\lim_{T \rightarrow \infty} g_3(T,\overline{\Delta})=\langle g_3(\overline{\Delta})\rangle$, {\it i.e.} every ring tends to travel at the same average speed (see Fig.~\ref{fig:Het}(A), $f_p=0$).
Conversely, perturbed systems display heterogeneously-distributed trajectories which can be partitioned into two sub-populations having small and large displacements, respectively. The former reflect the above-mentioned presence of caged rings and we also notice examples of single rings with temporally-heterogeneous dynamics, alternating slow and fast diffusion (see Fig.~\ref{fig:Het}(A), $f_p=0.3$). These observations are in agreement with the concept of permanent or transient caging due to threading TCs.  

Constraints with diverging life-times have been shown to trigger non-ergodic behaviours~\cite{Massignan2014}. We quantify these deviations form ergodic diffusion through  the ergodicity breaking parameter~\cite{Jeon2016}
\begin{equation}\label{eq:ErgBreakParam}
EB(T) \equiv \frac{\left[\langle g_3(T,\overline{\Delta})^2 \rangle - \langle g_3(T,\overline{\Delta})\rangle^2\right]}{\langle g_3(T,\overline{\Delta})\rangle^2} \, ,
\end{equation}
which captures how fast the single-ring trajectories $g_3(T,\overline{\Delta})$ narrow around the mean $\langle g_3(\overline{\Delta})\rangle$. For standard diffusive solutions, $EB(T) \sim T^{-1}$~\cite{MetzlerEB2014,Jeon2016} whereas non-ergodic systems display $EB(T) \sim T^0$~\cite{Deng2009}. As shown in Fig.~\ref{fig:Het}(C) (see also SM Figs.~S\ref{fig:AgeingN250}-S\ref{fig:AgeingN500})
ergodicity breaking can indeed be triggered by random pinning. Remarkably, even unperturbed ($f_p=0$) solutions of rings with $N=500$ and monomer density $\rho=0.4 \sigma^{-3}$ (Fig.~\ref{fig:Het}(D)) display slower convergence to ergodic behaviour,
thereby suggesting non-standard statistics in the waiting (sojourn) times of diffusing rings~\cite{Deng2009,MetzlerEB2014}.

Taken together, these results assemble into a picture where pinned rings induce a form of ``quenched disorder'', turning into ``traps'' with very large life-times~\cite{Massignan2014}, for the other rings.
To our knowledge, this is the first instance that spontaneous caging and deviations from standard ergodic behaviour is directly observed in unperturbed solutions of polymers (of any topology). 

\begin{figure}[t!]
	\centering
	\includegraphics[width=0.48\textwidth]{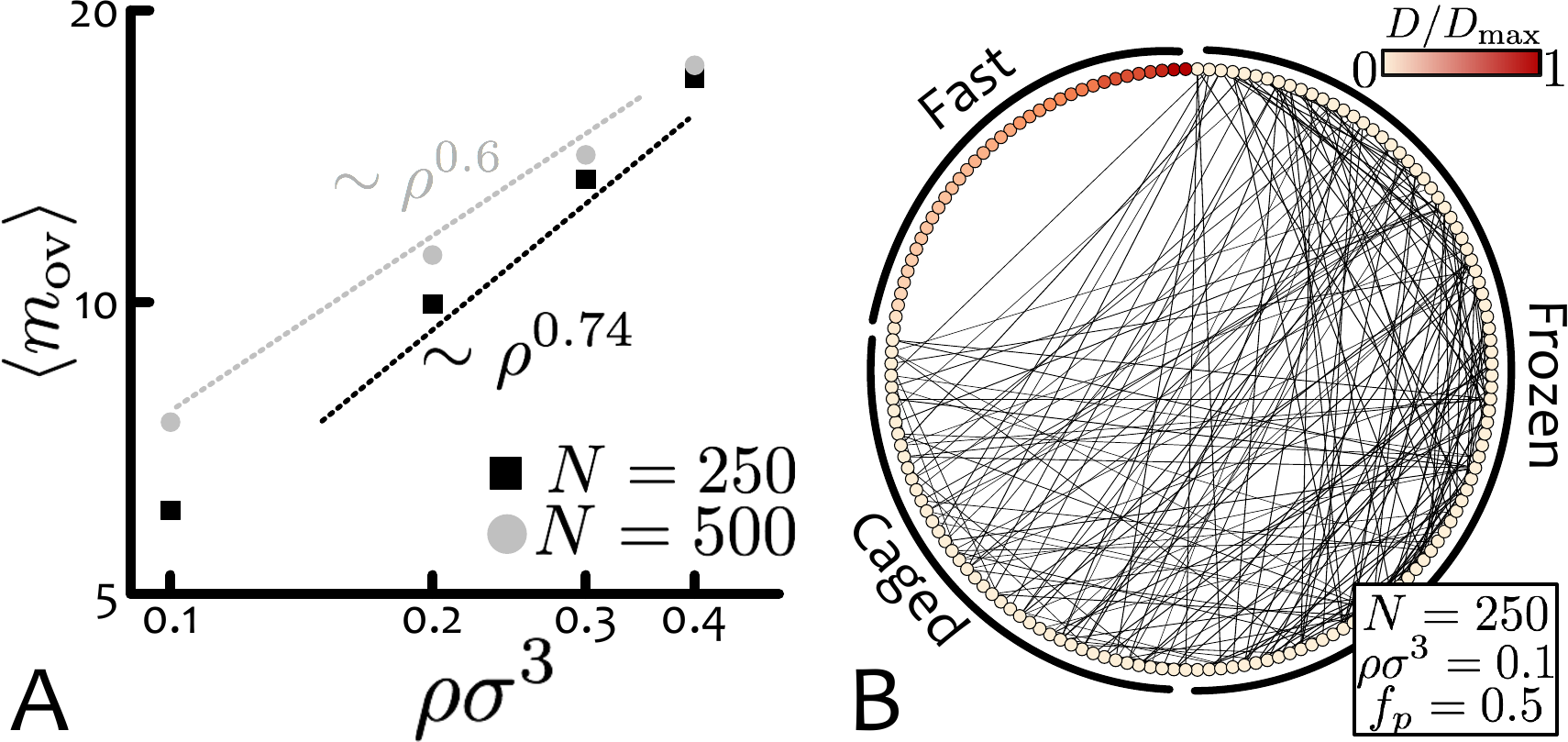}
	\caption{
		\textbf{Slowing Down of Overlapping Rings.}
		(\textbf{A})
		Average number of overlapping chains per ring, $\langle m_{\mathrm{ov}} (\rho) \rangle$.	 Dashed lines correspond to power-law behaviors determined from best fits to the data. Plotted values are listed in Table~S\ref{tab:CagedFractionPinnedRings} in SM.
		(\textbf{B})
		Abstract network representation for rings solutions:
		nodes (which represent rings) are colour coded according to corresponding diffusion coefficients, $D \equiv \lim_{\Delta \rightarrow \infty}g_3(\Delta)/6\Delta$. Edges between nodes are drawn if their weight is larger than 0.5, for clarity. Slow-moving rings overlap with frozen ones, whereas fast rings show little or no persistent overlap.
	}
	\label{fig:SurfaceMonomersContactingChains_1}
\end{figure}

Having investigated the heterogeneous dynamics of single rings, we conclude by connecting the observed non-Gaussian behaviour to the spatial organisation of the chains. 
By analogy with~\cite{Cates1986}, one may argue that a ring of size $R_g$ experiences an entropic penalty proportional to the average number of overlapping neighbours $\langle m_{\rm ov}\rangle$
\begin{equation}\label{eq:mov}
\frac{S}{k_B T} \sim \langle m_{\rm ov} \rangle \sim \frac{\rho}{N} \, R_g^3 \sim \rho^{\alpha} \, ,
\end{equation}
where we assume that~\cite{DeguchiJCP2009,Halverson2011_1,RosaEveraersPRL2014}, in the large-$N$ limit, the number of chains in a volume $R_g^3$ converges to a (density dependent) constant characterized by an exponent $\alpha<1$~\cite{NahaliRosa2016}, {\it i.e.} $R_g^3/N \sim \rho^{-(1-\alpha)}$.
In Eq.~\eqref{eq:mov}, $\langle m_{\rm ov} \rangle$ is defined as the average number of chains whose centres of mass are within $2 R_g$ from the centre of mass of any other ring and we find that $\langle m_{\rm ov}\rangle$ achieves a $N$-independent value with $\alpha \simeq 0.60-0.74$ (see Fig.~\ref{fig:SurfaceMonomersContactingChains_1}(A)).
Importantly, Eq.~\eqref{eq:mov} implies that {\it higher} monomer densities lead to a larger number of overlapping neighbours~\cite{NahaliRosa2016} and, in turn, larger entropic penalties~\cite{Cates1986}, which consequently drive more compact conformations.
On the other hand, results from Figs.~\ref{fig:g3sRho03}-\ref{fig:phasediag_scaling} suggest that denser systems are more susceptible to random pinning, and display glassy behaviour at {\it lower} values of $f_p$. 

This apparent contradiction can be reconciled by resorting to the following picture.
We model rings as nodes of an abstract network,
and a link between any two nodes indicates that the two corresponding rings overlap for a total time longer than half of the overall simulation runtime.
An example of such a network is given in Fig.~\ref{fig:SurfaceMonomersContactingChains_1}(B),
where nodes have been ordered and coloured according to the corresponding single-ring diffusion coefficients, $D \equiv \lim_{\Delta \rightarrow \infty}g_3(\Delta)/6\Delta$.
This representation intuitively shows that slow rings are connected (overlap) either with other slow rings or with frozen ones.
On the other hand, mobile rings ({\it i.e.}, rings with large diffusion coefficient) have virtually zero degree.
This representation thus directly connects static and dynamic features of rings in solution and, in particular, indicates that overlapping rings, arguably exerting TCs on one another, slow down the respective motion. 

To obtain then a quantitative estimation of how TCs affect the dynamics of rings,
we approximate the network as a Bethe lattice~\cite{RubinsteinColby} of coordination ({\it i.e.}, number of neighbours per node) $\langle m_{\rm ov} \rangle$.
Due to the hierarchical nature of the network, the maximum number of shells, $\bar{g}$, is given by
\begin{equation}\label{eq:MaxPathLength}
\bar{g} = \frac{\log \left( \frac{\langle m_{\rm ov} \rangle-2}{\langle m_{\rm ov} \rangle}(M-1) + 1 \right)}{\log(\langle m_{\rm ov} \rangle-1)} \, ,
\end{equation}
where $M$ is the total number of nodes (rings) of the network.
We now assume that the effect of pinning a single ring results in the caging of its first neighbours with an {\it unknown} probability $p_c$, of its second neighbours with probability $p_c^2$, and so on. The whole process therefore results in a ``caging cascade'' producing a fraction of trapped rings equal to
\begin{equation}\label{eq:CagedRingsPerSinglePinnedRing}
f^\prime_c = p_c \langle m_{\rm ov} \rangle \dfrac{\left(p_c \left( \langle m_{\rm ov} \rangle-1\right)\right)^{\bar{g}} -1 }{p_c \left( \langle m_{\rm ov} \rangle -1 \right) -1 } \, .
\end{equation}
For small $f_p$, all pinned rings may be assumed to act independently on their neighbours.
Then, the total fraction of caged rings, $f_c$, is
\begin{equation}\label{eq:CagedRingFractionPerPinnedRingFraction}
f_c = f_p \, f^\prime_c \, .
\end{equation}
Interestingly, Eqs.~(\ref{eq:CagedRingsPerSinglePinnedRing})-(\ref{eq:CagedRingFractionPerPinnedRingFraction}) link a measurable quantity (fraction of caged rings, $f_c$) to an imposed quantity (fraction of pinned rings, $f_p$) and, by inversion, allows to determine the caging (or threading) probability between close-by rings, $p_c$~\cite{MichielettoTurner2014,MichielettoTurner2016}. In particular, Eq.~(\ref{eq:CagedRingsPerSinglePinnedRing}) implies that the system becomes ``critical'' when $p_c = p_c^\dagger \equiv 1/(\langle m_{\mathrm{ov}} \rangle - 1)$,
for there exists a finite fraction $f_c$ of caged rings even in the limit $f_p \rightarrow 0$.

By combining Eqs.~(\ref{eq:CagedRingsPerSinglePinnedRing})-(\ref{eq:CagedRingFractionPerPinnedRingFraction}) and evaluating $f_c$ at $f_p=0.3$ as the rings displaying a single-ring diffusion coefficient $D/D_{\rm max} \simeq 0$ (see SM Fig.~S\ref{fig:PDeff}), we can numerically extract values for $p_c$ at any given $\rho$ (see Table~S\ref{tab:CagedFractionPinnedRings}). 
Interestingly, $p_c$ increases with $\rho$ up to where $p_c$ is approximately given by the predicted $p_c^\dagger$. Although we employ a crude approximation, we find that, curiously, the only two cases for which $p_c>p_c^\dagger$ are the ones displaying spontaneous ($f_p=0$) deviations from Gaussian behaviour (Fig.~\ref{fig:Pdisplacements}(D), $N=500$ and $\rho \sigma^3 \geq 0.3$).

\paragraph*{Conclusions --}

In this work, we have shown that dense solutions of semi-flexible ring polymers display rich, non-Gaussian behaviours under random pinning perturbations. 
Glassiness is observed at pinned fractions $f_p$ larger than a ``critical'' value $f_p^\dagger(\rho,N)$,
which obeys an empirical dependence on $\rho$ similar to the one previously reported for $N$~\cite{MichielettoTurner2016} (Fig.~\ref{fig:phasediag_scaling}).
As a consequence, we obtained novel, generic, and surprisingly simple, scaling relations for the threshold $\rho_g(N)$ and $N_g(\rho)$ marking the onset of spontaneous topological vitrification (Eqs.~\ref{eq:scaling}).

We have reported the first evidence of strong ergodicity breaking in solutions of rings, triggered for any $f_p>0$ and non-trivial convergence towards ergodicity has also been found for unperturbed solutions at high density (Fig.~\ref{fig:Het}).
These results can be rationalized by assuming that random pinning turn transient threading topological constraints into quenched disorder and permanent cages.
Overall, ring solutions appear to cluster into sub-systems with slow/fast diffusivities corresponding to large/little overlaps with other slow or pinned rings (Fig.~\ref{fig:SurfaceMonomersContactingChains_1}).

An intriguing finding of our work is that, even in the limit $f_p \rightarrow 0$, solutions of rings
may deviate from standard Gaussian behaviour (Figs.~\ref{fig:Pdisplacements}(D)-\ref{fig:Het}(D)) and turn into ``topological glasses'' provided $\rho$ or $N$ are large enough (Fig.~\ref{fig:phasediag_scaling}).
We have concluded that a topological glass may form when the probability $p_c$ of any pinned ring to cage any of its neighbours is $\geq p_c^\dagger$,
with $p_c^\dagger$ given by a simple analytical expression for networks in the Bethe lattice approximation.

We argue that the experimentally observed~\cite{Bras2014} non-Gaussian behaviour of ring polymers melts may be well reconciled with this picture.
At the same time, we hope that the present work will pave the way for future experiments and computer simulations.

\paragraph*{Acknowledgments --}
The authors thank C. Micheletti and T. Sakaue for insightful remarks on the manuscript and D. Levis for interesting suggestions.

\vspace*{-0.4 cm}
\bibliography{biblio,rings}

\begin{thebibliography}{48}
\bibitem[{Pha()}]{PhaseDiagFigNote}
\bibinfo{note}{We assume that the dependence on system density $\rho$ is solely
  included in $N_g$ and that $f_N$ is a constant parameter. Arguably, both will
  depend on microscopic details such as chain stiffness; nonetheless we expect
  the functional form to hold in general.}
\bibitem[{Mas()}]{MasterCurveNote}
\bibinfo{note}{This is because $f_N \log{\left[N/N_g(\rho)\right]}=f_\rho
  \log{\left[\rho/\rho_g(N)\right]}$ which leads to $\rho_g(N)=\rho
  N_g(\rho)^\eta N^{-\eta}$. Imposing that this equation must be valid at any
  $\rho$ one obtains that $\rho N_g(\rho)^\eta = const$ or $N_g(\rho)\sim
  \rho^{-1/\eta}$.}
\bibitem[{Not()}]{NoteOnPDeltaX}
\bibinfo{note}{The same results reported for $\Delta x$ are obtained for
  $\Delta y$, $\Delta z$ and $\Delta r$.}
\end{thebibliography}

\clearpage

\setcounter{section}{0}
\setcounter{figure}{0}
\setcounter{table}{0}
\setcounter{equation}{0}

\renewcommand{\figurename}{Fig. S}
\renewcommand{\tablename}{Table S}

{\large \bf Supplemental Material}

\tableofcontents

\clearpage

\section{Model and methods}\label{sec:ModMethods}

\subsection{The model}\label{sec:PolymerModel}

Solutions of ring polymers are modelled by resorting to the Kremer and Grest~\cite{KremerGrestJCP1990} polymer model.

Excluded volume interactions between beads (including consecutive ones along the contour of the chains) are described by the shifted and truncated Lennard-Jones (LJ) potential:
\begin{equation}\label{eq:LJ}
U_{\mathrm{LJ}}(r) = \left\{
\begin{array}{lr}
4 \epsilon \left[ \left(\frac{\sigma}{r}\right)^{12} - \left(\frac{\sigma}{r}\right)^6 + \frac14 \right] & \, r \le r_c \\
0 & \, r > r_c
\end{array} \right. \, ,
\end{equation}
where $r$ denotes the separation between the bead centers.
The cutoff distance $r_c=2^{1/6}\sigma$ is chosen so that  only the repulsion part of the Lennard-Jones is used.
The energy scale is set by $\epsilon=k_BT$ and the length scale by $\sigma$.
In the course of the paper, we adopt conventional LJ units with $\epsilon=1$ and $\sigma=1$.

Nearest-neighbour monomers along the rings contour length are connected through the finitely extensible nonlinear elastic (FENE) potential:
\begin{equation}\label{eq:Ufene}
U_{\mathrm{FENE}}(r) = \left\{
\begin{array}{lcl}
-0.5kR_0^2 \ln\left(1-(r / R_0)^2\right) & \ r\le R_0 \\ \infty & \
r> R_0 &
\end{array} \right. \, ,
\end{equation}
where $k = 30\epsilon/\sigma^2$ is the spring constant and $R_{0}=1.5\sigma$ is the maximum extension of the elastic FENE bond.

In order to maximize mutual chain interpenetration at relatively moderate chain length~\cite{mullerPRE2000} and hence reduce the computational effort,
we have introduced an additional bending energy penalty between
consecutive triplets of neighbouring beads along the chain in order to control polymer stiffness:
\begin{equation}\label{eq:Ubend}
U_{\mathrm{bend}}(\theta) = k_\theta \left(1-\cos \theta \strut\right) \, .
\end{equation}
Here, $\theta$
is the angle formed between adjacent bonds and $k_\theta = 5 \, k_B T$ is the bending constant.
With this choice, the polymer is equivalent to a worm-like chain with Kuhn length $l_K$ equal to $10 \sigma$~\cite{AuhlJCP2003}.

\subsection{Simulation details}\label{sec:SimDetails}
We consider equilibrated polymer solutions consisting of $M = 160$ and $M=80$ ring polymers made of $N=250$ and $N=500$ beads each, respectively.
The total number of monomers of each system is then fixed to $40'000$ monomer units.
We study solutions at four monomer densities, $\rho \sigma^3 = 0.1, 0.2, 0.3$ and $0.4$.
In particular, systems of rings with $\rho\sigma^3=0.1$ have been at the center of the previous study~\cite{MichielettoTurner2016},
and will be then used as a validation of the present work.

The static and kinetic properties of the chains are studied using fixed-volume and constant-temperature 
molecular dynamics (MD) simulations (NVT ensemble) with implicit solvent and periodic boundary conditions.
MD simulations are performed using the LAMMPS engine~\cite{lammps}.
The equations of motion are integrated using a velocity Verlet algorithm, in which all beads are weakly coupled to a Langevin heat bath 
with a local damping constant $\Gamma = 0.5 \, \tau_{\mathrm{LJ}}^{-1}$ where $\tau_{\mathrm{LJ}} = \sigma(m / \epsilon)^{1/2}$ is the Lennard-Jones
time and $m=1$ is the conventional mass unit for monomer and colloid particles.
The integration time step is set to $\Delta \tau = 0.012 \, \tau_{\mathrm{LJ}}$.

\subsection{Preparation of the initial configuration and system equilibration}\label{sec:IniConfigEquil}
Equilibrated solutions of ring polymers are prepared as described in Ref.~\cite{NahaliRosa2016}.
To avoid unwanted linking between close by rings, the chains were initially arranged inside a large simulation box at very dilute conditions.
In order to reach the correct monomer density of $\rho\sigma^3=0.1$ we performed then a short ($\approx 400 \, \tau_{\mathrm{LJ}}$) MD simulation
under fixed external pressure which shrinks the simulation box until it reaches the desired value.
Similarly, the other densities were reached by compressing the solutions under even higher imposed pressures.

Once any given system was prepared at the correct density, we switched to the NVT ensemble.
Then, each system was equilibrated by performing single MD runs up to
$1 \cdot 10^9 \Delta \tau = 12 \cdot 10^6 \tau_{\mathrm{LJ}}$ (for $N=250$) and $2 \cdot 10^9 \Delta \tau = 24 \cdot 10^6 \tau_{\mathrm{LJ}}$ (for $N=500$),
during which the center of mass of each chain moves on average a distance comparable to $\approx 3-4$ times its corresponding gyration radius, $R_g$
(for more details, see the Supplementary Material of Ref.~\cite{NahaliRosa2016}).

\subsection{Molecular dynamics runs}\label{sec:MDRuns}
After equilibration,
ring dynamics was studied by performing MD simulations up to $1 \cdot 10^9 \Delta \tau = 12 \cdot 10^6 \tau_{\mathrm{LJ}}$ for both $N=250$ and $N=500$.
We studied systems with different pinning fractions $f_p$ of the total number of rings in the range $f_p=0.1-0.7$.
For comparison we have also considered unperturbed rings solutions, corresponding to pinning fraction $f_p=0$.

\subsection{Calculation of diffusion coefficients}\label{sec:DiffCoeffsCalc}
The ring asymptotic diffusion coefficient at given ($N$, $\rho$, $f_p$) is defined as:
$$D(\rho, f_p) \equiv \lim_{\Delta\rightarrow\infty} \frac{\langle g_3(\Delta) \rangle} {6\Delta} \, ,$$
where $\langle g_3(\Delta) \rangle$ is the mean-square displacement of the chain center of mass.
By following the same procedure employed in Ref.~\cite{MichielettoTurner2016},
$D(\rho, f_p)$ is computed by standard best fit of the long-time behaviour of the corresponding $\langle g_3(\Delta) \rangle$ to a linear function.
This is done regardless the long-time behavior of $\langle g_3(\Delta) \rangle$ is effectively linear or saturates to a plateau like in frozen set-up's.
In the latter case, the evaluated $D(\rho, f_p)$ should be then best considered as an upper value to the ``true'' asymptotic behavior.

\clearpage

\section{Supplemental Figures}

%
\begin{figure*}[t!]
	\centering
	\includegraphics[width=0.95\textwidth]{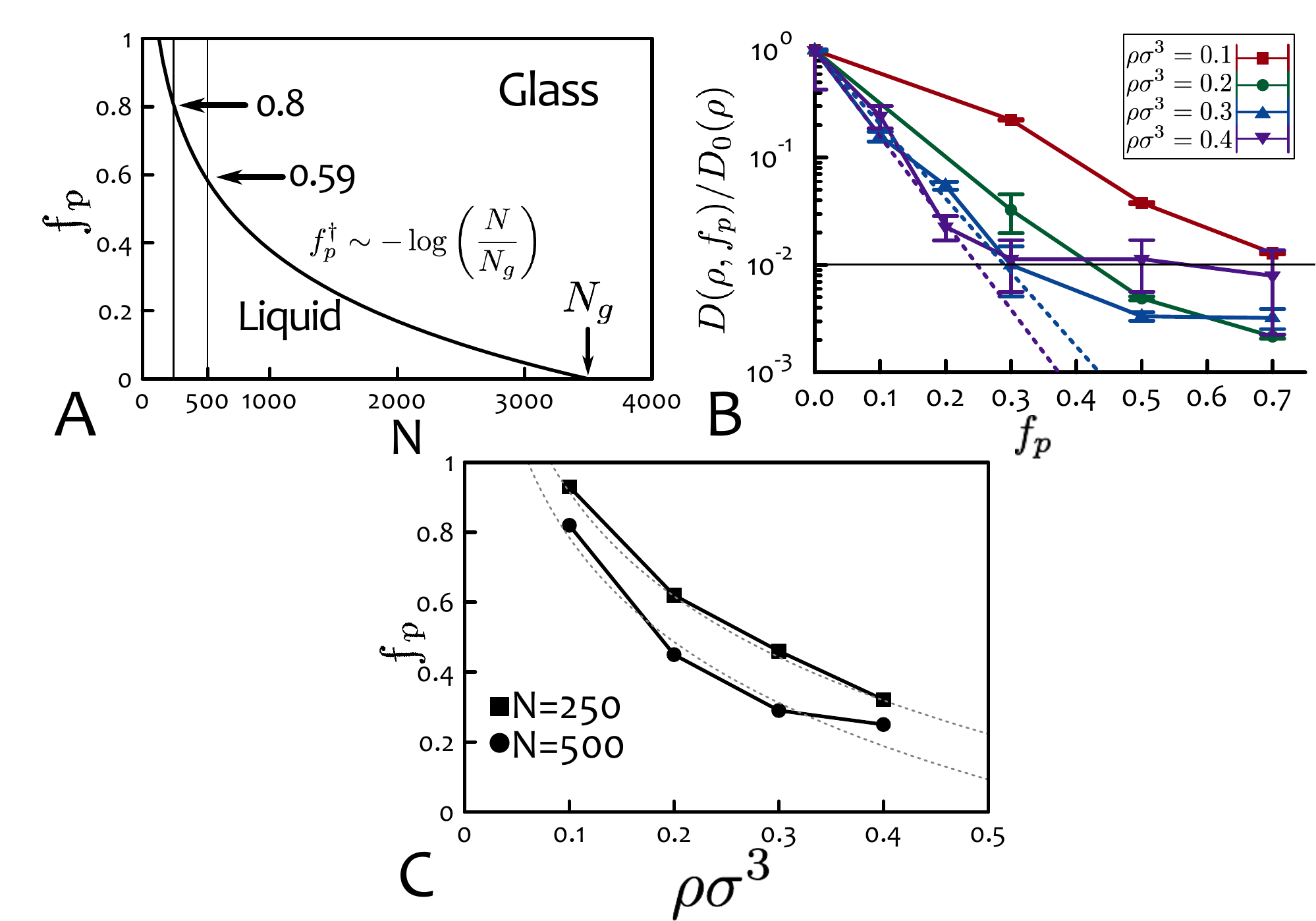}
	\caption{
	{\bf Phase Diagrams and Scaling.}
	(\textbf{A})
	Phase diagram for solutions of semi-flexible ring polymers at monomer density $\rho=0.1\sigma^{-3}$~\cite{MichielettoTurner2016}.
	The two vertical lines represent the values of $N$ chosen in this work and the intersections $f_p^\dagger = -f_N\log(N/N_g)$ are the predicted values of $f_p^\dagger$ for the onset of glassiness at $\rho=0.1\sigma^{-3}$.
	$N_g$ is the (empirical) value at which the system is expected to vitrify at zero pinning fraction.
	(\textbf{B})
	Scaled diffusion coefficient $D(\rho,f_p)/D_0(\rho)$ ($D_0(\rho) \equiv \lim_{\Delta \rightarrow \infty}\langle g_3(\Delta) \rangle/6\Delta$ with $f_p=0$) as a function of $f_p$ for $N=500$. The plots shows the exponential decay of the diffusion coefficient. For high densities and large $f_p$, the reported values are overestimates due the insufficient length of simulation runs. We therefore extract the exponential decay from the values of $D(\rho,f_p)$ measured at small $f_p$.      
	(\textbf{C})
	Phase diagram in the plane $(f_p,\rho\sigma^3)$ for the ring solutions studied in this work.
	The data points are obtained by fitting $D(\rho,f_p)/D_0(\rho)$ with an exponential function $d(f_p)=\exp{(-f_p/a)}$ and by solving $d(f_p)=0.01$.
	This gives the ``critical'' $f_p^\dagger$ at fixed $\rho$ and $N$.
	The functional dependence appears to follow a scaling relation similar to that found for $N$, {\it i.e.} $f_p^\dagger(\rho)=-f_\rho\log{\rho/\rho_g(N)}$ where $\rho_g(N=250)=0.84$ and $\rho_g(N=500)=0.6$ are the theoretical threshold densities for the spontaneous onset of glassiness.
	Both curves have $f_\rho=0.43$ suggesting that this parameter depends very weakly on $N$ or $\rho$.
	As a consequence, the data points collapse onto a master curve by plotting $f_p^\dagger(x=\rho/\rho_g(N))/f_\rho=-\log(x)$ (see Fig.~\ref{fig:phasediag_scaling}(B), main text).  
	}
	\label{fig:phasediag}
\end{figure*}
\begin{figure*}[t!]
	\centering
	\includegraphics[width=0.9\textwidth]{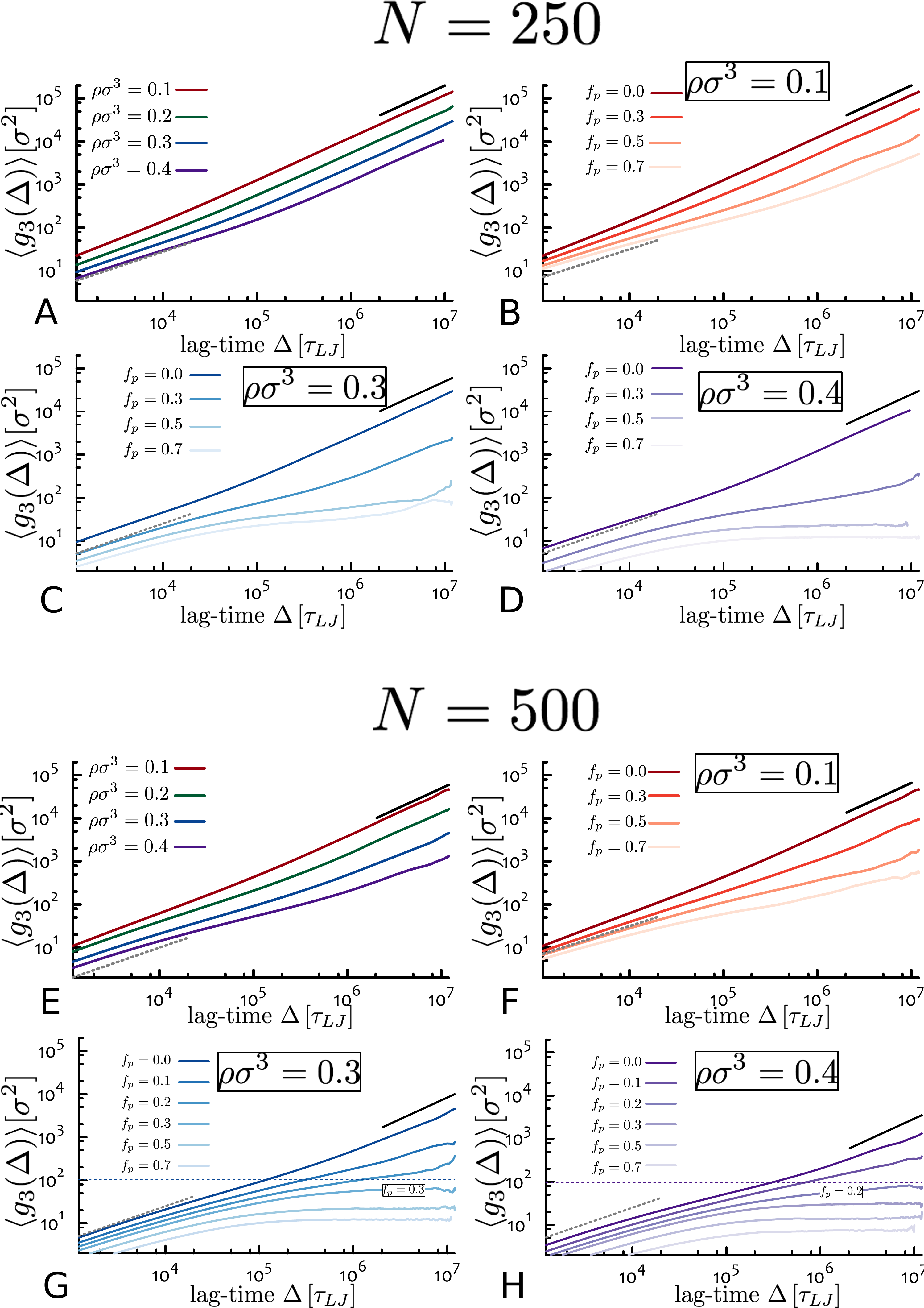}
	\caption{
	{\bf Dynamics of Ring Polymers in Solutions of Density $\rho$ and Ring Pinning Fraction $f_p$.}
	The curves correspond to the mean-square displacement of ring center of mass, $\langle g_3(\Delta) \rangle$ (see main text for details), at lag-time $\Delta$, and for chain sizes $N=250$ and $N=500$.
	Solid black lines correspond to the long-time diffusive ($\sim \Delta^1$) regime, whereas dashed grey lines represent the short-time sub-diffusive ($\sim \Delta^{3/4}$) regime~\cite{Halverson2011_2,Bras2014,GrosbergSoftMatter2014}.
	Horizontal dashed lines in panels G and H are for the mean-square ring diameter, $4 \langle R_g^2\rangle$.
	}
	\label{fig:g3sRhoOthers}
\end{figure*}
\begin{figure*}[t!]
	\centering
	\includegraphics[width=1\textwidth]{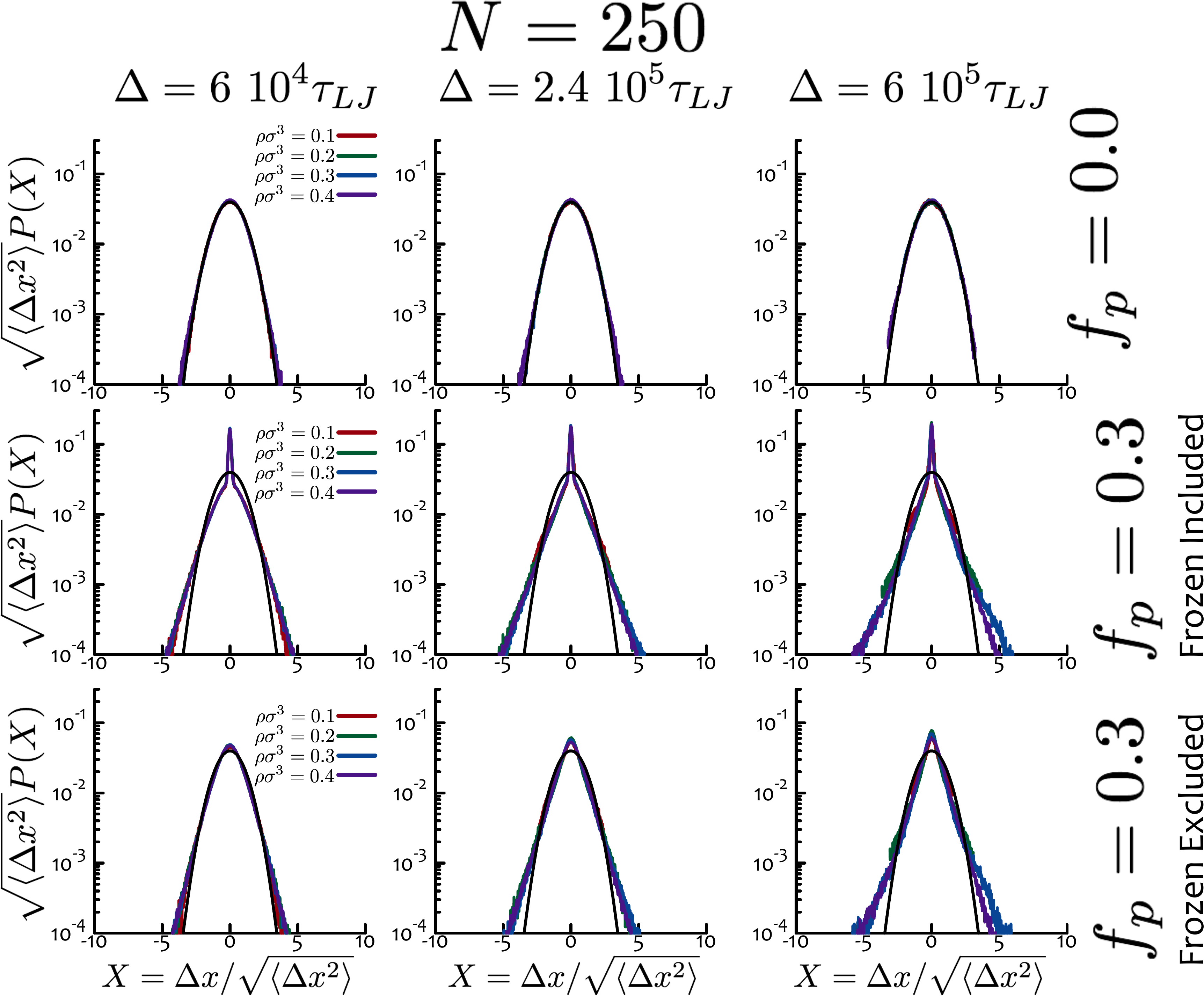}
	\caption{
	{\bf Probability Distribution Functions of $1d$ Displacements.}
	The figure shows a collection of $P(\Delta x)$ for $N=250$ and $f_p=0$ (top row) and $f_p=0.3$ (middle and bottom rows).
	The curves are rescaled by the average displacement, {\it i.e.} $X= \Delta x /\sqrt{\langle \Delta x^2 \rangle}$ in order to better visualise deviations from standard Gaussian behaviour (represented by the solid black line).
	Middle and bottom rows show $P(X)$ for all rings or only the non-pinned ones, respectively.
	One can observe that at $f_p=0.3$ the Gaussian behaviour observed for $f_p=0$ is not recovered, and that pinned rings are represented by a narrow spike.
	Once pinned rings are removed from the measurement,
	the distributions display the non-Gaussian dynamics of the non-pinned rings,
	alternating caging (higher $P(X)$ at small $X$ with respect to standard Gaussian) and sudden jumps (exponential tails at large $X$).
	For the broadest overview, each column represents different lag-times $\Delta$ at which corresponding $P(X)$'s are measured.   
	}
	\label{fig:PDisplacementsA}
\end{figure*}
\begin{figure*}[t!]
	\centering
	\includegraphics[width=0.9\textwidth]{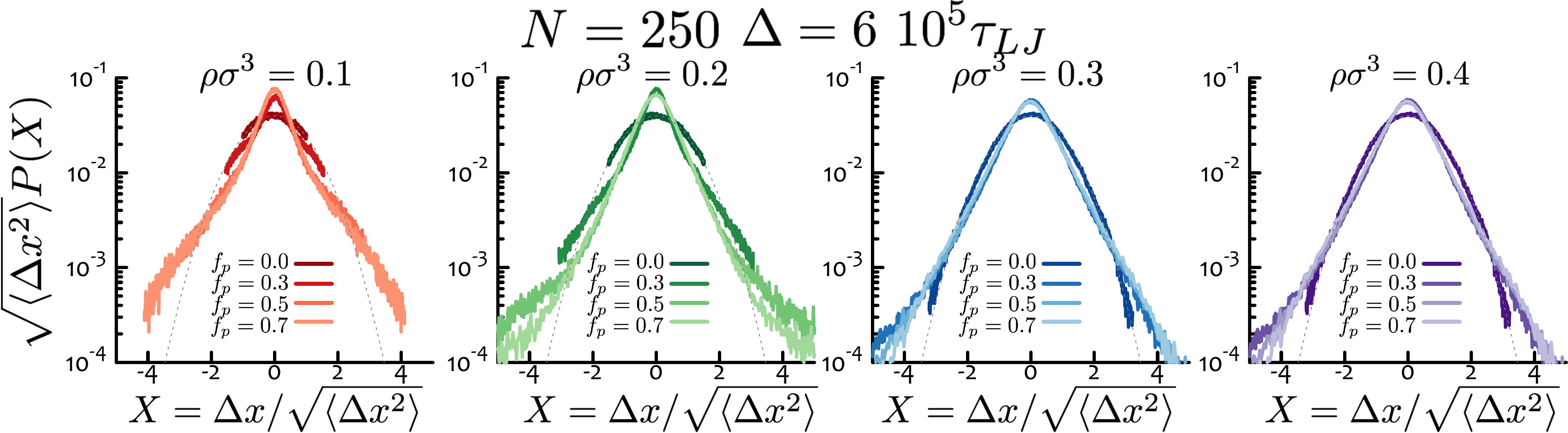}
	\caption{
		{\bf Probability Distribution Functions of $1d$ Displacements.}
		$P(X)$ (as in Fig.~S\ref{fig:PDisplacementsA}) for $N=250$, $\Delta=6$ $10^5 \tau_{LJ}$ and different densities.
		In the plots, different values of $f_p$ are compared.
		Increasing values of $f_p$ trigger more marked deviations from the Gaussian behaviour (dashed line).  
	}
	\label{fig:PDisplacementsB}
\end{figure*}
\begin{figure*}[t!]
	\centering
	\includegraphics[width=0.95\textwidth]{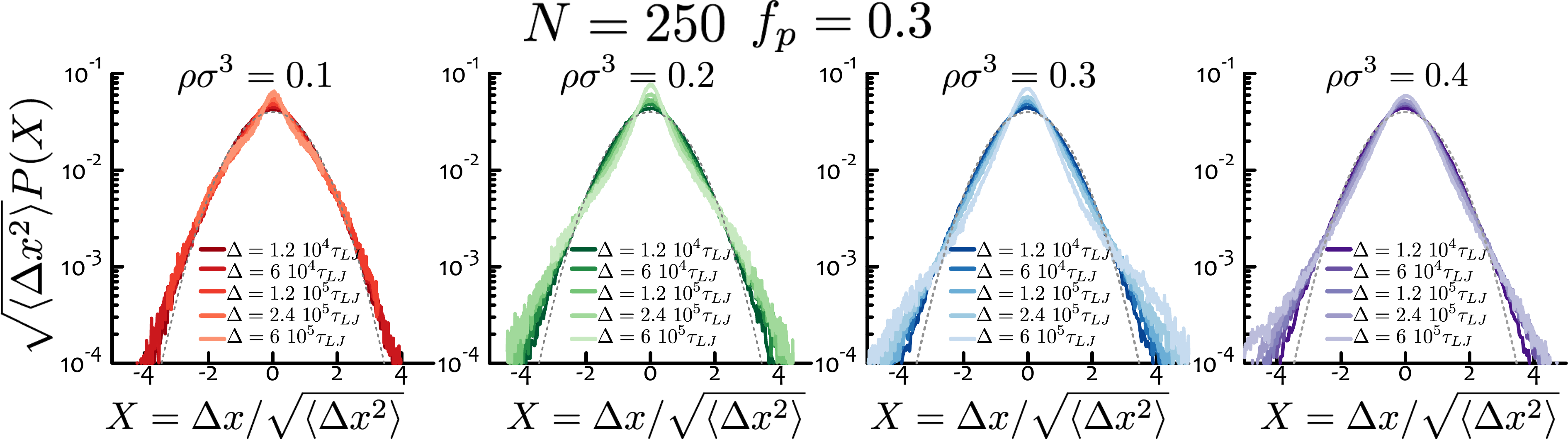}
	\caption{
			{\bf Probability Distribution Functions of $1d$ Displacements.}
			$P(X)$ (as in Fig.~\ref{fig:PDisplacementsA}) for $N=250$, $f_p=0.3$, different densities and increasing values of $\Delta$ (from dark to light colours).
			Larger values of $\Delta$ enhance deviations from the Gaussian behaviour (dashed line).
			In particular, the central peak (which captures the caged rings) becomes more pronounced once the non-caged ones are able to diffuse farther than the mean $\sqrt{\langle \Delta x^2 \rangle}$, thereby fattening the tails at large $X$.
	}
	\label{fig:PDisplacementsC}
\end{figure*}
\begin{figure*}[t!]
	\centering
	\includegraphics[width=0.9\textwidth]{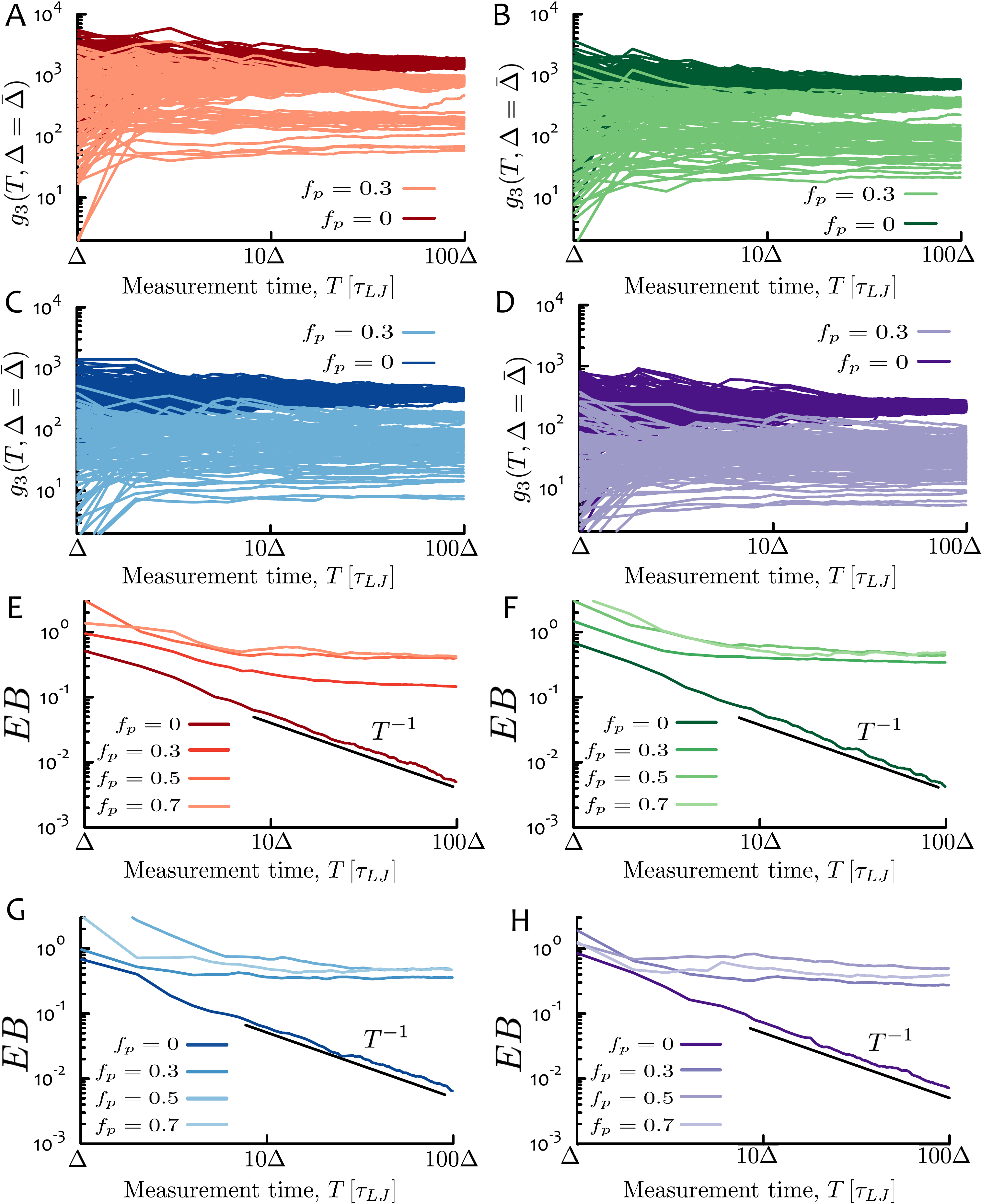}
	\caption{
	{\bf Time-Averaged Displacement of Rings in Solution ($N=250$).}
	(A-D)
	Curves, $g_3 = g_3(T, \Delta=\overline{\Delta})$ {\it vs.} measurement time $T$ at fixed lag-time $\Delta=\overline{\Delta}=1.2\times10^5$ $\tau_{\mathrm{LJ}}$.
	(E-H)
	Ergodicity-breaking (EB) parameter defined as~\cite{MetzlerEB2014,Jeon2016}:
	$EB = EB(T) \equiv \left[\langle g_3(T,\Delta=\overline{\Delta})^2 \rangle - \langle g_3(T,\Delta=\overline{\Delta})\rangle^2\right]/\langle g_3(T,\Delta=\overline{\Delta})\rangle^2$.
	The heterogeneity in $g_3$ decreases with measurement time as $T^{-1}$ for unperturbed systems, as expected for standard diffusion.
	On the other hand, for perturbed ($f_p>0$ systems), EB flattens and the system displays heterogeneous dynamics and ergodicity breaking.
	}
	\label{fig:AgeingN250}
\end{figure*}
\begin{figure*}[t!]
	\centering
	\includegraphics[width=0.9\textwidth]{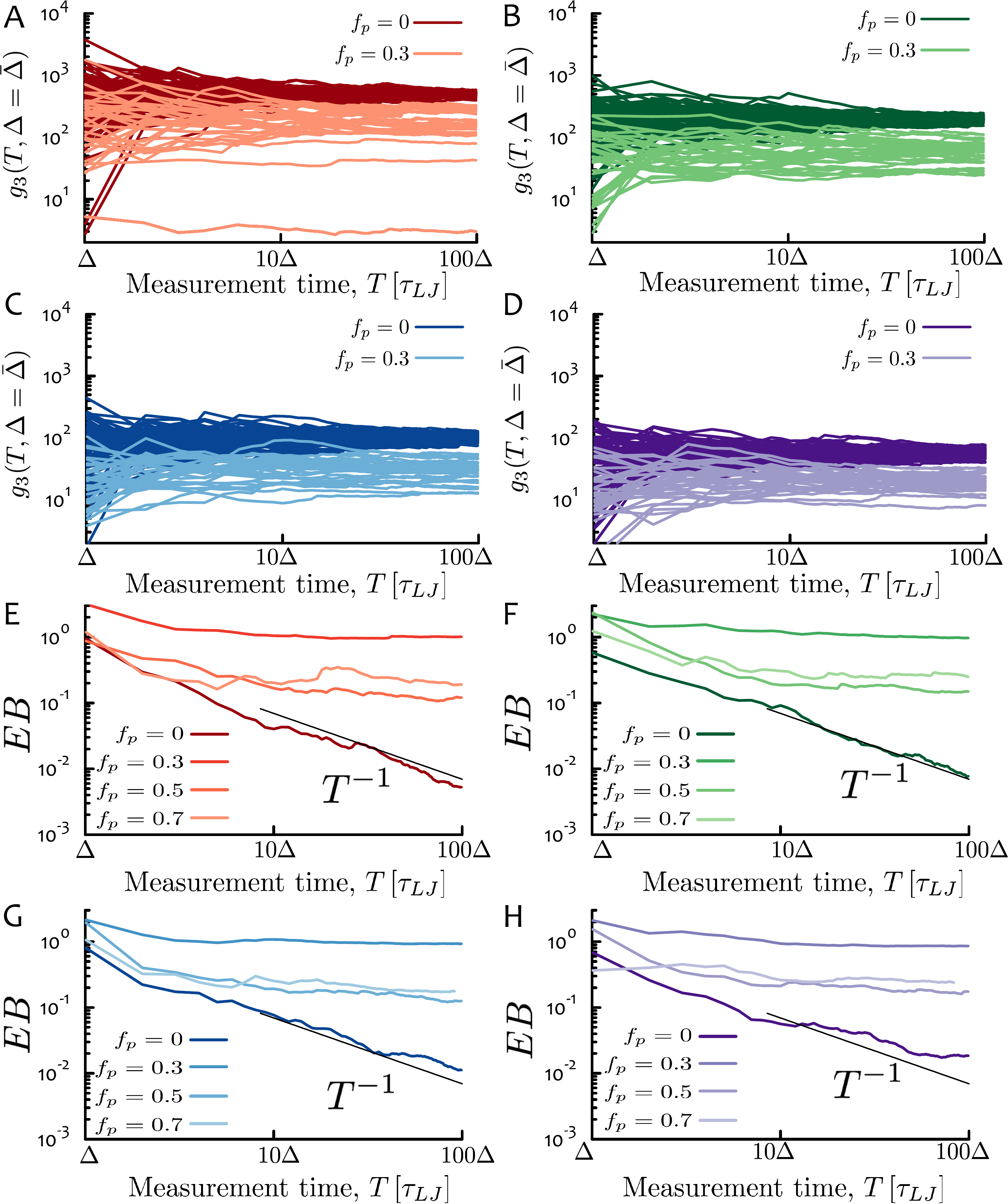}
	\caption{
	{\bf Time-Averaged Displacement of Rings in Solution ($N=500$).}
	Notation, symbols and color code are as in Fig.~S\ref{fig:AgeingN250}.
	}
	\label{fig:AgeingN500}
\end{figure*}
\begin{figure*}[t!]
	\centering
	\includegraphics[width=0.7\textwidth]{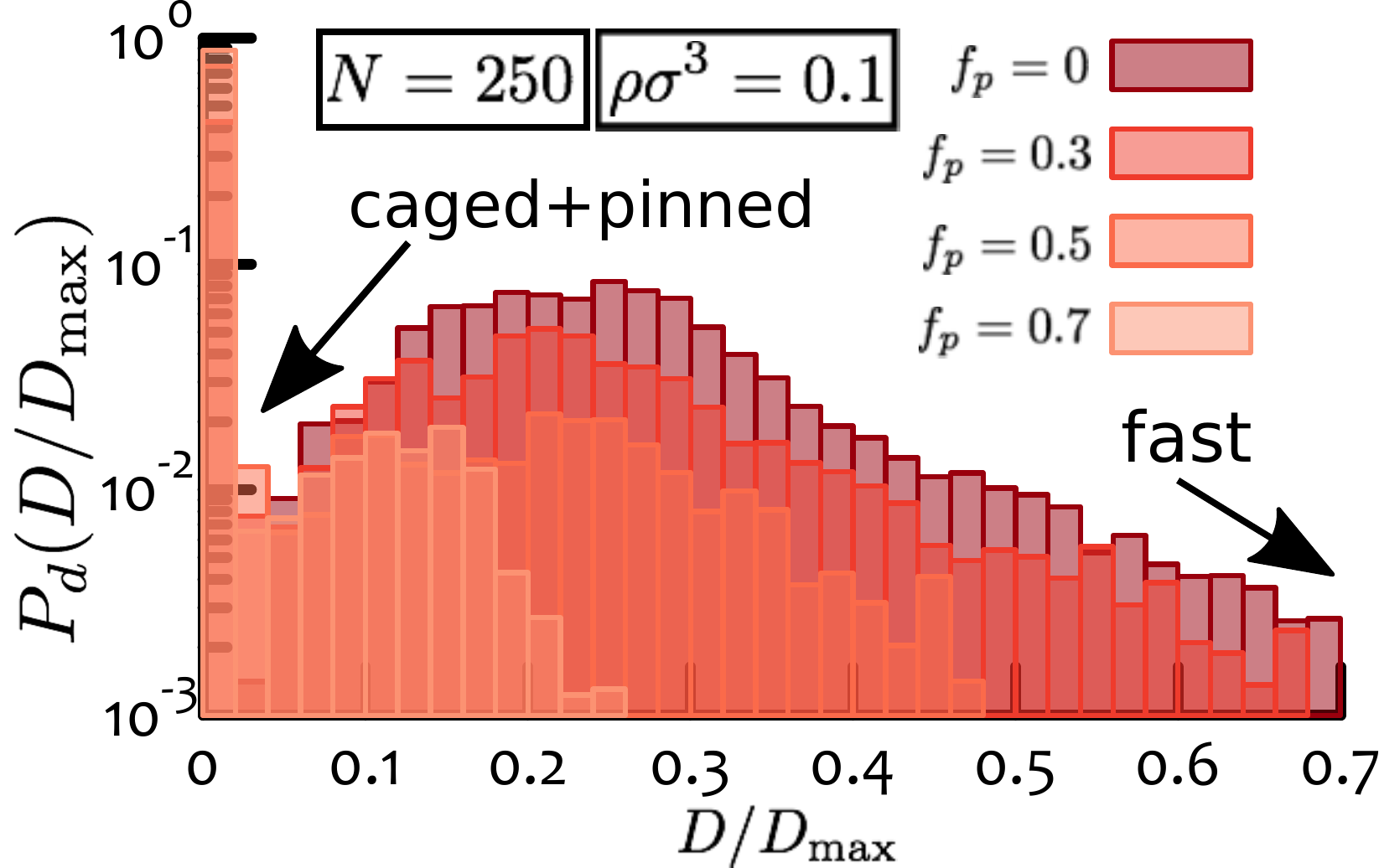}
	\caption{\textbf{Distribution of Diffusion Coefficients.}
	Example of distribution function, $P_d(D/D_{\rm max})$, of the scaled rings diffusion coefficients $D/D_{\rm max}$ for $N=250$, $\rho\sigma^3=0.1$ and different values of $f_p$.
	The first bin contains both, pinned and caged rings; from this, we can readily extract the fraction of caged rings as $f_c=P_d(0)-f_p$.
	The values obtained for the systems studied in this work are reported in table  ~S\ref{tab:CagedFractionPinnedRings}.
	}
	\label{fig:PDeff}
\end{figure*}

\clearpage

\section{Supplemental Tables}

%
\begin{table*}
\begin{tabular}{|c|c|c|c|c|c|c|c|c|c|c|}
\hline
& \multicolumn{5}{c|}{$N=250$} & \multicolumn{5}{c|}{$N=500$} \\
\hline
$\rho\sigma^3$ & $\langle m_{\mathrm{ov}} \rangle$ & $\bar g$ & $f_c(f_p=0.3)$ & $p_c^\dagger \equiv 1/(\langle m_{\mathrm{ov}} \rangle - 1)$ & $p_c$ & $\langle m_{\mathrm{ov}} \rangle$ & $\bar g$ & $f_c(f_p=0.3)$ & $p_c^\dagger \equiv 1/(\langle m_{\mathrm{ov}} \rangle - 1)$ & $p_c$\\
\hline
$0.1$ & $6.098$ & $2.874$ & $0.129$ & $0.196$ & $0.053$ & $7.520$ & $2.175$ & $0.283$ & $0.153$ & $0.080$\\
$0.2$ & $9.958$ & $2.213$ & $0.312$ & $0.112$ & $0.063$ & $11.184$ & $1.804$ & $0.556$ & $0.098$ & $0.094$\\
$0.3$ & $13.383$ & $1.953$ & $0.515$ & $0.081$ & $0.070$ & $14.186$ & $1.641$ & $0.623$ & $0.076$ & $0.086$\\
$0.4$ & $17.032$ & $1.785$ & $0.606$ & $0.062$& $0.065$ & $17.549$ & $1.519$ & $0.649$ & $0.060$ & $0.076$\\
\hline
\end{tabular}
\caption{
\label{tab:CagedFractionPinnedRings}
Measured values for:
(1)
the average number of overlapping chains per ring, $\langle m_{\mathrm{ov}} \rangle$;
(2)
the maximum number of shells in the Bethe-lattice representation of rings solutions, $\bar{g}$, Eq.~(\ref{eq:MaxPathLength}) in main text;
(3)
fraction of caged rings, $f_c$, at pinning fraction $f_p=0.3$ (value chosen for corresponding to the smallest pinning fraction used in this work);
(4)
``critical'' caging probability, $p_c^\dagger \equiv 1/(\langle m_{\mathrm{ov}} \rangle - 1)$,
corresponding to a finite fraction $f_c$ of caged rings in the limit $f_p \rightarrow 0$;
(5)
caging probability, $p_c$, obtained from Eqs.~(\ref{eq:CagedRingsPerSinglePinnedRing})-(\ref{eq:CagedRingFractionPerPinnedRingFraction}) in main text.
}
\end{table*}

\end{document}